\documentclass[runningheads]{llncs}
\usepackage[T1]{fontenc}
\usepackage{amsmath}
\usepackage{amssymb}
\usepackage{color}
\usepackage{tabularray}
\usepackage{graphicx}
\begin{document}
\title{Data Augmentation for Environmental Sound Classification Using Diffusion Probabilistic Model with Top-k Selection Discriminator }
\titlerunning{DPMs for Audio Data Augmentation}
%
%
\author{Yunhao Chen\inst{1}\orcidID{0000-0002-8134-2314} \and
Yunjie Zhu\inst{2}\and
Zihui Yan\inst{1}\and 
Jianlu Shen\inst{1}
Zhen Ren\inst{1}
Yifan Huang\inst{1}
}
\authorrunning{Y. Chen et al.}
%
\institute{Jiangnan University, Wuxi, 214000, China \\ 
\email{1191200221@stu.jiangnan.edu.cn}
\and
University of Leeds, Leeds, LS2 9JT, United Kingdom
\\
}
\maketitle              
\begin{abstract}        
Despite consistent advancement in powerful deep learning techniques in recent years, large amounts of training data are still necessary for the models to avoid overfitting. Synthetic datasets using generative adversarial networks (GAN) have recently been generated to overcome this problem. Nevertheless, despite advancements, GAN-based methods are usually hard to train or fail to generate high-quality data samples. In this paper, we propose an environmental sound classification augmentation technique based on the diffusion probabilistic model with DPM-Solver$++$ for fast sampling. In addition, to ensure the quality of the generated spectrograms, we propose a  top-k selection technique to filter out the low-quality synthetic data samples. According to the experiment results, the synthetic data samples have similar features to the original dataset and can significantly increase the classification accuracy of different state-of-the-art models compared with traditional data augmentation techniques. The public code is available on \url{https://github.com/JNAIC/DPMs-for-Audio-Data-Augmentation}.
\keywords {diffusion probabilistic models, data augmentation, environmental sound classification}
\end{abstract}
\section{Introduction}
In recent years, deep learning models have been advancing in the sound classification task like the convolutional neural networks (CNN)\cite{b2_traditional_augmentation}, \cite{b6_CNN}, \cite{b7_CNN2}, transformer-based network \cite{b3_AST}, \cite{b4_HTS}, CNN-RNN network\cite{b5_CNN_RNN}. However, these methods are hungry for considerable amounts of data for efficient performance due to their large amounts of parameters. Consequently, the main challenge the deep learning methods are confronted with is the limited samples of the training dataset. The most challenging aspect of developing deep learning supervised models is data annotation. It is a labour-intensive process that is expensive and time-consuming, especially when the deep learning methods require a lot of samples. Consequently, data augmentation is proposed to overcome the limited data samples problem. 

Traditional data augmentation for audio classification includes flip, rotation, scale, crop, translation, noise, pitch shifting, masking, etc.\cite{b8_tranditional_augmentation,b2_traditional_augmentation} However, these methods are based on simple or linear transformations. Compared with the complexity of deep learning-based methods, these augmentations can not efficiently improve the performance of CNN-based or transformer-based classifier methods. As a result, to enhance the accuracy of deep learning-based methods, it is necessary to either apply data augmentation techniques that involve transformations of similar complexity to the deep learning models, or employ a model that can represent the probability distribution of the real data samples. Accordingly, researchers take advantage of the generative models, especially the generative adversarial network (GAN) \cite{b8_tranditional_augmentation,b2_traditional_augmentation,b5_CNN_RNN,b9_gan_aug2,b10_gan_aug3} to synthesize new data samples with qualities similar to real data samples. 

However, despite the considerable amounts of GANs' applications in data augmentation, they are subject to unstable training processes, model collapse issues and failure to represent a broad enough data distribution. \cite{b11_gan_dis1,b12_gan_dis2,b13_gan_dis3,b14_gan_dis4} Consequently, the GANs are challenging to be scaled rapidly to a new application. 

 On the other hand, diffusion probabilistic models \cite{b15_diffusion1,b16_diffusion2} are becoming much more popular than GANs. Moreover, the diffusion-based generative models have been proven that they can beat the GANs on many tasks such as image synthesis \cite{b17_diffusion_good}, medical image synthesis where the data samples are limited\cite{b18_diffusion_good2}, topology optimization\cite{b19_diffusion_good3} and so on. Therefore, we adopt diffusion probabilistic models for high-quality data augmentation. However, the popular diffusion models' sampling procedures are mainly based on denoising diffusion implicit models (DDIM) \cite{b21_DDIM},  which require 50 to 100 steps to generate high-quality data samples. This is extremely time-consuming when generating thousands of data samples. On the other hand, DPM-Solver$++$ \cite{b22_DPM} only requires 10 to 20 steps to generate similar results compared with the DDIM method.
 Consequently, we employ the DPM-Solver$++$ for our sampling strategy. 
 
  Diffusion probabilistic models can generate diverse data samples from complex distributions by reversing a Markov chain of Gaussian diffusion processes. However, the quality of the data samples generated by these models is not always satisfactory, as they may contain artefacts, blurriness or inconsistency with the target distribution. Consequently, we propose a Top-k Selection method to filter out inappropriate data samples based on a pretrained model to address this problem. Our method can improve the quality of data generation without modifying the diffusion probabilistic models.

 This paper aims to improve the environmental sound classification process using a conditional diffusion probabilistic network framework for high-quality data augmentation with a discriminator and DPM-Solver$++$. To summarize, the main contributions of this
paper are as follows:
 
 1) We present the first study on applying conditional diffusion probabilistic network frameworks  to generate high-quality data samples for environmental sound classification tasks based on a popular sound dataset, UrbanSound8K \cite{b23_us8k}.

2)We propose a post-processing approach called top-k selection 
  based on a pre-trained discriminator. This approach automatically eliminates samples with low quality and insufficient representation after training.

3) We evaluate seven state-of-the-art deep learning (DL) models for environmental sound classification on the dataset with real and synthetic data samples generated by data augmentation. We train the models from scratch, without transfer learning, and show significant accuracy improvement with synthetic data.

\section{Method}
This section explains the methods for diffusion probabilistic models, DPM-Solver$++$ and data augmentation.

\subsection{Diffusion Probabilistic Models}

Diffusion probabilistic models(DPMs)\cite{b_24_diffusion}\cite{b_25_diffusion} are a class of generative models that convert Gaussian noise into samples from a learned data distribution via an iterative denoising process. Non-equilibrium thermodynamics serves as the basis for diffusion models. To gradually introduce random noise to the data, \cite{b_24_diffusion}\cite{b_25_diffusion} establish a Markov chain of diffusion steps. Then they figure out how to reverse the diffusion process to create the desired data samples from the noise by virtual deep learning methods, the details of which will be discussed next.

\subsubsection{Forward Diffusion Process}

A forward diffusion process is defined as a process that adds Gaussian noise to a data sample $\mathbf{x}_{0}$ sampled from a real data distribution $q(\mathbf{x})$ over $T$ steps, resulting in a sequence of noisy samples $\mathbf{x}_{1}, \ldots, \mathbf{x}_{T}$. The amount of noise added at each step is determined by a variance schedule $\left ({\beta_{t} \in(0,1)}_{t=1}^{T} \right) $.

\begin{equation} \label{forward} 
q\left(\mathbf{x}_{t} \mid \mathbf{x}_{t-1}\right)=\mathcal{N}\left(\mathbf{x}_{t} ; \sqrt{1-\beta_{t}} \mathbf{x}_{t-1}, \beta_{t} \mathbf{I}\right) \quad q\left(\mathbf{x}_{1: T} \mid \mathbf{x}_{0}\right)=\prod_{t=1}^{T} q\left(\mathbf{x}_{t} \mid \mathbf{x}_{t-1}\right) \end{equation}

The data sample $\mathbf{x}_{0}$ gradually loses its distinctive features as $t$ increases. When $T$ approaches infinity, $\mathbf{x}_T$ converges to an isotropic Gaussian distribution.

An advantage of this process is that we can obtain samples at any arbitrary time step using a closed-form expression with the reparameterization trick. 

$
\begin{array}{rlr}
\mathbf{x}_{t} & =\sqrt{\alpha_{t}} \mathbf{x}_{t-1}+\sqrt{1-\alpha_{t}} \boldsymbol{\epsilon}_{t-1} &   \\
\\
& =\sqrt{\alpha_{t} \alpha_{t-1}} \mathbf{x}_{t-2}+\sqrt{1-\alpha_{t} \alpha_{t-1}} \overline{\boldsymbol{\epsilon}}_{t-2} &  \\
\\
& =\cdots \\
\\
& =\sqrt{\bar{\alpha}_{t}} \mathbf{x}_{0}+\sqrt{1-\bar{\alpha}_{t}} \boldsymbol{\epsilon} 

\end{array}
$

We can come to the following equation:

\begin{equation}
\label{forward2}
    q\left(\mathbf{x}_{t} \mid \mathbf{x}_{0}\right)  =\mathcal{N}\left(\mathbf{x}_{t} ; \sqrt{\bar{\alpha}_{t}} \mathbf{x}_{0},\left(1-\bar{\alpha}_{t}\right) \mathbf{I}\right)
\end{equation}

where $ \alpha_{t}=1-\beta_{t} $, 
 $\bar{\alpha}_{t}=\prod_{i=1}^{t} \alpha_{i} $,  $\boldsymbol{\epsilon}_{t-1}, \boldsymbol{\epsilon}_{t-2}, \cdots \sim \mathcal{N}(\mathbf{0}, \mathbf{I})$ and $\overline{\boldsymbol{\epsilon}}_{t-2} $ merges\ two Gaussians.

\subsubsection{Reverse Diffusion Process}

 Reversing this process and sampling from \\ $q\left(\mathbf{x}_{t-1} \mid \mathbf{x}_{t}\right)$ would enable us to reconstruct the true sample from a Gaussian noise input, $\mathbf{x}_{T} \sim \mathcal{N}(\mathbf{0}, \mathbf{I})$. If $\beta_{t}$ is sufficiently small,  $q\left(\mathbf{x}_{t-1} \mid \mathbf{x}_{t}\right)$ will also be Gaussian. However, estimating $q\left(\mathbf{x}_{t-1} \mid \mathbf{x}_{t}\right)$ is not feasible because it requires using the entire dataset. Therefore, we need to learn a model $p_{\theta}$ that approximates these conditional probabilities for running the reverse diffusion process.

 \begin{equation}
 \label{reverse2}
p_{\theta}\left(\mathbf{x}_{t-1} \mid \mathbf{x}_{t}\right)=\mathcal{N}\left(\mathbf{x}_{t-1} ; \boldsymbol{\mu}_{\theta}\left(\mathbf{x}_{t}, t\right), \mathbf{\Sigma}_{\theta}\left(\mathbf{x}_{t}, t\right)\right)   
\end{equation}

Where $\theta$ is a learnable parameter vector in the Gaussian distribution’s mean function $\boldsymbol{\mu}_{\theta}\left(\mathbf{x}_{t}, t\right)$ and standard deviation function $\mathbf{\Sigma}_{\theta}\left(\mathbf{x}_{t},t\right)$, the data samples generated by this distribution can be represented as:

 \begin{equation}
\label{reverse1}
 p_{\theta}\left(\mathbf{x}_{0: T}\right)=p\left(\mathbf{x}_{T}\right) \prod_{t=1}^{T}  p_{\theta}\left(\mathbf{x}_{t-1} \mid \mathbf{x}_{t}\right) \quad \\ 
 \end{equation}


By applying the learned parameters $\theta$ to the mean function $\boldsymbol{\mu}_{\theta}\left(\mathbf{x}_{t}, t\right)$ and the standard deviation function $\mathbf{\Sigma}_{\theta}\left(\mathbf{x}_{t}, t\right)$. In brief, the forward diffusion process adds noise to the data sample, while the reverse diffusion process removes the noise and creates new data samples.

\subsubsection{Training Objective of Diffusion Probabilistic Models}

Similar to Variational Autoencoder \cite{b_36_vae}, the variational lower bound can be used to optimize the negative log-likelihood as follows:

\vspace{0.5cm}
$
\begin{array}{rlr}

-\log p_{\theta}\left(\mathbf{x}_{0}\right) & \leq-\log p_{\theta}\left(\mathbf{x}_{0}\right)+D_{\mathrm{KL}}\left(q\left(\mathbf{x}_{1: T} \mid \mathbf{x}_{0}\right) \| p_{\theta}\left(\mathbf{x}_{1: T} \mid \mathbf{x}_{0}\right)\right) \\
\\
& =-\log p_{\theta}\left(\mathbf{x}_{0}\right)+\mathbb{E}_{\mathbf{x}_{1:T} \sim q\left(\mathbf{x}_{1:T} \mid \mathbf{x}_{0}\right)}\left[\log \frac{q\left(\mathbf{x}_{1: T} \mid \mathbf{x}_{0}\right)}{p_{\theta}\left(\mathbf{x}_{0: T}\right) / p_{\theta}\left(\mathbf{x}_{0}\right)}\right] \\
\\
& =-\log p_{\theta}\left(\mathbf{x}_{0}\right)+\mathbb{E}_{q}\left[\log \frac{q\left(\mathbf{x}_{1: T} \mid \mathbf{x}_{0}\right)}{p_{\theta}\left(\mathbf{x}_{0: T}\right)}+\log p_{\theta}\left(\mathbf{x}_{0}\right)\right] \\
\\
& =\mathbb{E}_{q}\left[\log \frac{q\left(\mathbf{x}_{1: T} \mid \mathbf{x}_{0}\right)}{p_{\theta}\left(\mathbf{x}_{0: T}\right)}\right] \\

\end{array}
$
\vspace{0.5cm}

We can come to the following equations:

\begin{equation}\label{trainging}
 L_{\mathrm{VLB}}  =\mathbb{E}_{q(\mathbf{x}_{0:T})}\left[\log \frac{q\left(\mathbf{x}_{1: T} \mid \mathbf{x}_{0}\right)}{p_{\theta}\left(\mathbf{x}_{0: T}\right)}\right] \geq-\mathbb{E}_{q(\mathbf{x} 0)} \log p_{\theta}\left(\mathbf{x}_{0}\right)
 \end{equation}

To make each term in the equation analytically computable, we can reformulate the objective as a combination of several terms involving KL divergence and entropy.  The objective can be rewritten as follows:

\vspace{0.5cm}
$
\begin{array}{l} 
L_{\mathrm{VLB}}=\mathbb{E}_{q\left(\mathbf{x}_{0: T}\right)}\left[\log \frac{q\left(\mathbf{x}_{1: T} \mid \mathbf{x}_{0}\right)}{p_{\theta}\left(\mathbf{x}_{0: T}\right)}\right] \\
\\
=\mathbb{E}_{q}\left[\log \frac{\prod_{t=1}^{T} q\left(\mathbf{x}_{t} \mid \mathbf{x}_{t-1}\right)}{p_{\theta}\left(\mathbf{x}_{T}\right) \prod_{t=1}^{T} p_{\theta}\left(\mathbf{x}_{t-1} \mid \mathbf{x}_{t}\right)}\right] \\
\\
=\mathbb{E}_{q}\left[-\log p_{\theta}\left(\mathbf{x}_{T}\right)+\sum_{t=1}^{T} \log \frac{q\left(\mathbf{x}_{t} \mid \mathbf{x}_{t-1}\right)}{p_{\theta}\left(\mathbf{x}_{t-1} \mid \mathbf{x}_{t}\right)}\right] \\
\\
=\mathbb{E}_{q}\left[-\log p_{\theta}\left(\mathbf{x}_{T}\right)+\sum_{t=2}^{T} \log \frac{q\left(\mathbf{x}_{t-1} \mid \mathbf{x}_{t}, \mathbf{x}_{0}\right)}{p_{\theta}\left(\mathbf{x}_{t-1} \mid \mathbf{x}_{t}\right)}+\log \frac{q\left(\mathbf{x}_{T} \mid \mathbf{x}_{0}\right)}{q\left(\mathbf{x}_{1} \mid \mathbf{x}_{0}\right)}+\log \frac{q\left(\mathbf{x}_{1} \mid \mathbf{x}_{0}\right)}{p_{\theta}\left(\mathbf{x}_{0} \mid \mathbf{x}_{1}\right)}\right] \\
\\
=\mathbb{E}_{q}\left[\log \frac{q\left(\mathbf{x}_{T} \mid \mathbf{x}_{0}\right)}{p_{\theta}\left(\mathbf{x}_{T}\right)}+\sum_{t=2}^{T} \log \frac{q\left(\mathbf{x}_{t-1} \mid \mathbf{x}_{t}, \mathbf{x}_{0}\right)}{p_{\theta}\left(\mathbf{x}_{t-1} \mid \mathbf{x}_{t}\right)}-\log p_{\theta}\left(\mathbf{x}_{0} \mid \mathbf{x}_{1}\right)\right] \\
\\
=\mathbb{E}_{q}[\underbrace{D_{\mathrm{KL}}\left(q\left(\mathbf{x}_{T} \mid \mathbf{x}_{0}\right) \| p_{\theta}\left(\mathbf{x}_{T}\right)\right)}_{L_{T}}+\sum_{t=2}^{T} \underbrace{D_{\mathrm{KL}}\left(q\left(\mathbf{x}_{t-1} \mid \mathbf{x}_{t}, \mathbf{x}_{0}\right) \| p_{\theta}\left(\mathbf{x}_{t-1} \mid \mathbf{x}_{t}\right)\right)}_{L_{t-1}} \\ \\ 
-\underbrace{\log p_{\theta}\left(\mathbf{x}_{0} \mid \mathbf{x}_{1}\right)}_{L_{0}}]

\end{array}
$

Since $x_{0}$ follows a fixed data distribution and $x_{T}$ is a Gaussian noise, $L_{T}$ is a constant term. We can also interpret $L_{0}$ as the entropy of the multivariate Gaussian distribution, because $p_{\theta}\left(x_{0} \mid x_{1}\right)$ is a Gaussian distribution with mean $\mu_{\theta}\left(x_{1}, 1\right)$ and covariance matrix $\mathbf{\Sigma}_{\theta}$.
The loss term  $L_{t}, t \in[1,2,3, \ldots, T-1]$  can be parameterized as:

\begin{equation}\label{objective}
L_{t}=E_{x_{0}, \epsilon_{t}}\left[\frac{\beta_{t}^{2}}{2 \alpha_{t}(1-\bar{\alpha_{t}}) \mathbf{\Sigma}_{\theta}^{2}}\left\|\epsilon_{t}-\epsilon_{\theta}\left(\sqrt{\bar{\alpha}_{t}} x_{0}+\sqrt{1-\bar{\alpha}_{t}} \epsilon_{t}, t\right)\right\|^{2}\right]+C
\end{equation}

According to \cite{b_25_diffusion}, the diffusion model can be trained more effectively by using a simplified objective that does not include the weighting term: 

\begin{equation}\label{objectivity2}
    L_{\text {simple }}(\theta)=E_{x_{0}, \epsilon_{t}}\left[\left\|\epsilon_{t}-\epsilon_{\theta}\left(\sqrt{\bar{\alpha}_{t}} x_{0}+\sqrt{1-\bar{\alpha}_{t}} \epsilon_{t}, t\right)\right\|^{2}\right]+C
\end{equation}

\subsubsection{Classifier-Free Guidance for Conditional Generation}

Conditional diffusion steps can be performed by combining the scores from both a conditional and an unconditional diffusion model \cite{b_37_classifier_guided}. The unconditional diffusion probabilistic model $p_{\theta}(\mathbf{x})$ is parameterized by a score estimator $\boldsymbol{\epsilon}_{\theta}\left(\mathbf{x}_{t}, t\right)$ , while the conditional model $p_{\theta}(\mathbf{x} \mid y)$ is parameterized by $\boldsymbol{\epsilon}_{\theta}\left(\mathbf{x}_{t}, t, y\right)$. A single neural network can learn these two models simultaneously.

The implicit classifier’s gradient can be expressed with conditional and unconditional score estimators. The classifier-guided modified score, which incorporates this gradient, does not depend on a separate classifier.

\vspace{0.5cm}

\begin{equation}\label{condition}
\nabla_{\mathbf{x}_{t}} \log p\left(y \mid \mathbf{x}_{t}\right)  =-\frac{1}{\sqrt{1-\bar{\alpha}_{t}}}\left(\boldsymbol{\epsilon}_{\theta}\left(\mathbf{x}_{t}, t, y\right)-\boldsymbol{\epsilon}_{\theta}\left(\mathbf{x}_{t}, t\right)\right) \\
\end{equation}
\begin{equation}\label{condition2}
\overline{\boldsymbol{\epsilon}}_{\theta}\left(\mathbf{x}_{t}, t, y\right) =(w+1) \boldsymbol{\epsilon}_{\theta}\left(\mathbf{x}_{t}, t, y\right)-w \boldsymbol{\epsilon}_{\theta}\left(\mathbf{x}_{t}, t\right)
\end{equation}

 In this study, we present the first study on applying the classifier-free guidance
diffusion probabilistic model to generate high-quality data samples for environmental sound classification tasks.

\subsection{DPM-Solver$++$ and DPM-Solver}

One of the main challenges of working with diffusion probabilistic models is the high computational cost and time required to generate data samples from the complex posterior distributions. To overcome this limitation, we adopt DPM-Solver$++$ as our sampling method.

 DPM-Solver is a high-order solver that can generate high-quality samples in around 10 steps by solving the diffusion ODE with a data prediction model. DPM-Solver++ is an improved version of DPM-Solver that can handle guided sampling by using thresholding methods to keep the solution matching the training data distribution.

The DPM-Solver$++$ and DPM-Solver raise the efficiency of training-free samplers to a new level to generate high-quality samples in the "few-step sampling" regime. This is the regime in which the sampling can be done within approximately 10 steps of sequential function evaluations. The DPM-Solver$++$ and DPM-Solver tackle the alternative problem of sampling from DPMs by solving the corresponding diffusion ordinary differential equations (ODEs) of DPMs. Moreover, diffusion ODEs have a semi-linear structure, meaning they comprise two parts: a linear function dependent on the data variable and a nonlinear function parameterized by neural networks. Consequently, the DPM-Solver$++$ and DPM-Solver use a precise formulation of the solutions of diffusion ODEs by analytically computing the linear portion of the solutions, thereby preventing the discretization mistake that the corresponding discretization would otherwise cause. In addition, it is possible to simplify the solutions to an exponentially weighted summation of the neural network by employing change-of-variable. This can be done efficiently. Such an integral is very special and can be efficiently approximated by numerical methods for exponential integrators. 

Customized solver for diffusion ODEs is shown below:

\begin{equation} \label{eq5}  
\boldsymbol{x}_{t_{i-1} \rightarrow t_{i}}=\frac{\alpha_{t_{i}}}{\alpha_{t_{i-1}}} \tilde{\boldsymbol{x}}_{t_{i-1}}-\alpha_{t_{i}} \sum_{n=0}^{k-1} \hat{\boldsymbol{\epsilon}}_{\theta}^{(n)}\left(\hat{\boldsymbol{x}}_{\lambda_{t_{i-1}}}, \lambda_{t_{i-1}}\right) \int_{\lambda_{t_{i-1}}}^{\lambda_{t_{i}}} e^{-\lambda} \frac{\left(\lambda-\lambda_{t_{i-1}}\right)^{n}}{n !} \mathrm{d} \lambda
\end{equation}

Where $\boldsymbol{x}_{s}$ is an initial value at time $s>0$,  
 $\boldsymbol{x}_{t}$ is the solution  at time $t \in[0, s]$ and $\lambda_{t}:=\log \left(\alpha_{t} / \Sigma_{t}\right)$. As  $\lambda(t)=\lambda_{t}$  is a strictly decreasing function of  $t$ , it has an inverse function  $t_{\lambda}(\cdot)$  satisfying  $t=t_{\lambda}(\lambda(t))$ . The DPM-Solver$++$ further changes the subscripts of  $\boldsymbol{x}$  and  $\boldsymbol{\epsilon}_{\theta}$  from  $t$  to  $\lambda$  and denote  $\hat{\boldsymbol{x}}_{\lambda}:=\boldsymbol{x}_{t_{\lambda}(\lambda)} ,  \hat{\boldsymbol{\epsilon}}_{\theta}\left(\hat{\boldsymbol{x}}_{\lambda}, \lambda\right):=\boldsymbol{\epsilon}_{\theta}\left(\boldsymbol{x}_{t_{\lambda}(\lambda)}, t_{\lambda}(\lambda)\right)$. The $\mathcal{O}\left(h_{i}^{k+1}\right)$ is omitted in the above equation because it is a high-order error. 

The main difference between DPM-Solver and DPM-Solver++ is that the former is designed for sampling without guidance, while the latter is designed for sampling with guidance. Sampling with guidance means using additional information, such as text or images, to guide the sampling process of DPMs, which can improve the sample quality and diversity.

 The primary technique that DPM-Solver++ uses to adapt to guided sampling is thresholding. Thresholding is a method to keep the solution of the diffusion ODE within the training data distribution by applying a hard or soft threshold to the pixel values or the latent variables. Thresholding can reduce the noise and ambiguity in the generated images and improve the sample quality and diversity. DPM-Solver++ adopts two types of thresholding: dynamic thresholding and static thresholding. Dynamic thresholding adjusts the threshold value according to the noise level and the guidance scale, while static thresholding uses a fixed threshold value for all steps. DPM-Solver++ combines both types of thresholding to balance the trade-off between stability and efficiency.

 \subsection{Data Augmentation }
 
Data augmentation is a potent tactic to broaden the current data range and enable model training without requiring new data collection. In this research, standard and intelligent data augmentation methodologies are also taken into consideration. Two distinct audio data deformations are applied in conventional data augmentation. 

First, certain background noises were added to the data samples, including crowd, street, and restaurant sounds (the background noises were taken from publicly available recordings made available on the "freesound.org" website \cite{b_26_free_sound}).
Second, the records are subjected to pitch shifting \cite{b_27_pitch_shifting}. The audio samples' pitch is adjusted by a half-octave (up and down) to produce various sounds. The audio stream is subjected to each contortion before being transformed into the input representation. Thirdly, the audio sample augmentation implements the time stretch \cite{b_31_time_stretch}.

For intelligent data augmentation, we use the U-net structure in the \cite{b_29_pytorch_diffusion} for diffusion probabilistic models and DPM-Solver$++$ for the sampling schedule.

\subsection{ Top-k Selection  Pretrained Discriminator }

One of the challenges of using DPMs for data augmentation is that the quality of the generated samples may vary depending on the amount of available data and computational resources. To address this issue, we propose to use a pretrained discriminator network to filter out the low-quality samples and retain only the ones that are realistic and diverse enough to augment the training data. The discriminator network is an Xception \cite{b_38_inception} trained on the entire dataset to classify the images into their corresponding labels. The filtering criterion is based on the top-k accuracy of the discriminator, i.e., we accept a generated sample if its accurate label is among the top-k predictions of the discriminator. Otherwise, we reject it. This way, we ensure that the generated samples are visually plausible and semantically consistent with their labels. The number of accepted samples can be expressed as follows:

\begin{equation} \label{eq6} G=\sum_{i=1}^N{\mathbb{I} \left( f_k\left( \boldsymbol{x}_i,c_i \right) =c_i \right)} \end{equation} where $c_i$ is the label of the i-th generated sample, $\boldsymbol{x}_i$ is the i-th generated sample, $f_k$ is the discriminator network with top-k prediction, and $N$ is the number of generation epochs.

\subsection{DL models for Environmental Sound Classification}

One of the main objectives of this study is to assess the quality of the synthetic data samples generated by DPMs for environmental classification tasks. To this end, we propose to use the synthetic data samples to augment the original training data, and then train different deep learning (DL) classifiers on the augmented data. We hypothesize that the augmented data can enhance the diversity and robustness of the training data, and thus improve the performance of the DL classifiers for weed recognition. To test this hypothesis, we select seven state-of-the-art DL models from different architectures and paradigms, namely ResNet-50 \cite{b_28_ResNet}, Xception \cite{b_38_inception}, ConViT-tiny \cite{b_39_convit}, mobilevitv2-50 \cite{b_40_mobile}, mobilevitv2-150 \cite{b_40_mobile}, ConvNext-tiny \cite{b_41_conv} and Deit III \cite{b_42_deit}. These models are implemented using the timm library \cite{b_32_timm}, and their hyperparameters are set to their default values as suggested by the authors. We evaluate the performance of these models on the UrbanSound8K dataset.

\section{Experiments}

\subsection{Experiments Pipeline}

\begin{figure}
\centering

\centerline{\includegraphics[width=1.0\columnwidth]{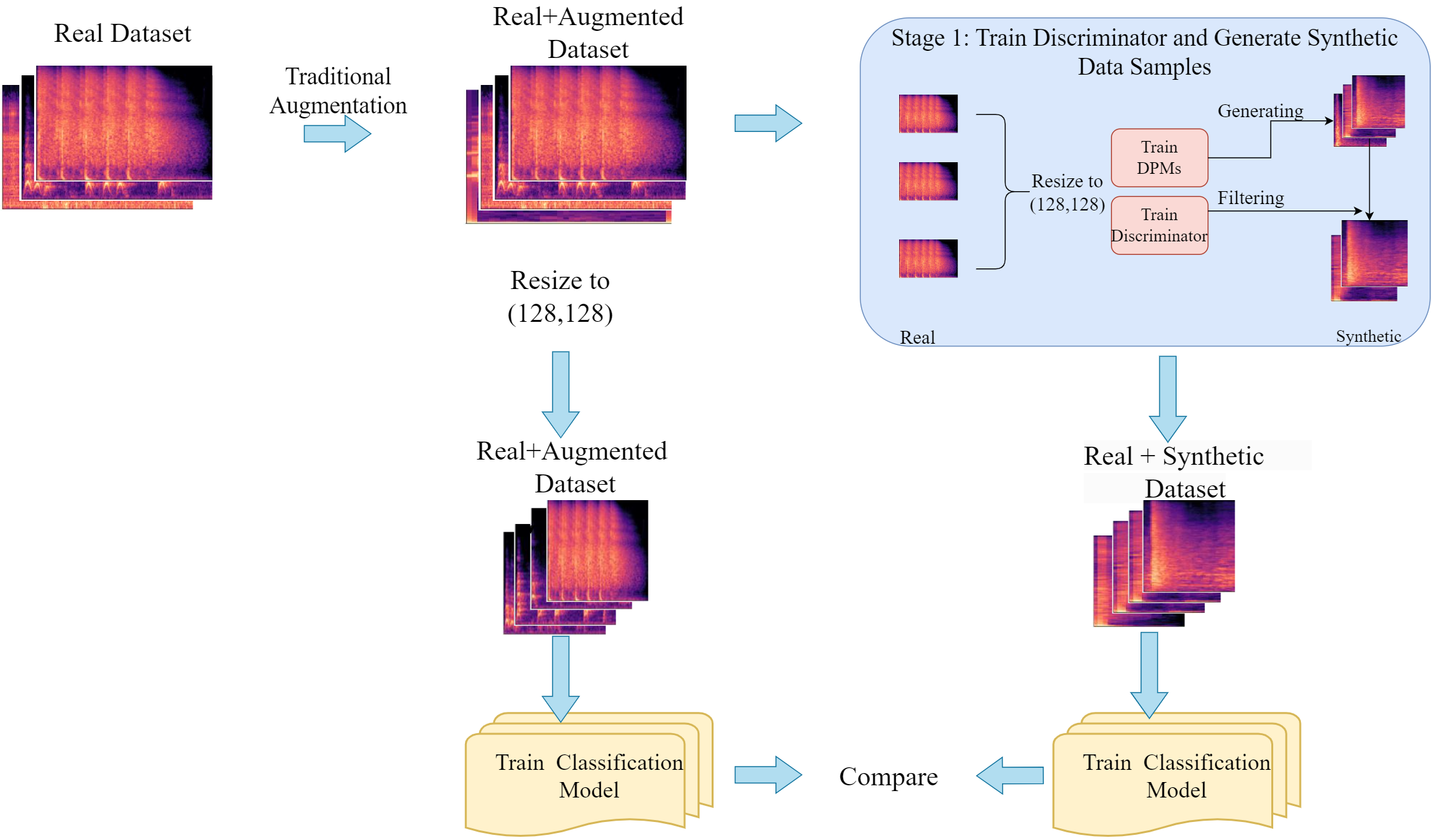}}
\caption{\quad An overview of the proposed pipeline for generating synthetic data samples. The pipeline consists of four main components: data augmentation, diffusion probabilistic modeling, discriminator filtering, and comparison experiments.
}
\end{figure}

The whole Pipeline is shown in Fig 1.

The procedure can be divided into several steps in the proposed experimental settings. Initially, the audio is transformed into mel-spectrogram, which serves as the foundation for the subsequent steps. Following this, the training samples are augmented and resized into 128$\times$128, and the resulting samples are incorporated into the original dataset. Simultaneously, a discriminator model is trained on the augmented dataset.

Once the augmented dataset is prepared, a diffusion probabilistic model is trained on it. This model is then employed to generate data samples. The discriminator model, which was previously trained, is used to filter out unqualified data samples from the generated samples. Finally, \textbf{8730 qualified synthetic data samples} are generated.

Lastly, two comparison experiments are conducted to evaluate the effectiveness of the models. The first experiment involves training a model using the original data samples and traditional augmentation techniques. In contrast, the second experiment trains a model using the original data samples along with synthetic data samples. 

\subsection{UrbanSound8K Dataset}


UrbanSound8K is an audio dataset that contains 8732 labelled sound excerpts($<=4s$) of urban sounds from 10 classes: air conditioner, car horn, children playing, dog bark, drilling, engine idling, gunshot, jackhammer, siren, and street music. The classes are drawn from the urban sound taxonomy. All excerpts are taken from field recordings uploaded to www.freesound.org.

The dataset can be used for various tasks such as urban sound classification, sound event detection, acoustic scene analysis, etc.

\subsection{Hyperparameters Setting}

\subsubsection{Data Preprocessing}\label{AA}
This section investigates the performance of the diffusion probabilistic model using the UrbanSound8K dataset to augment the data samples and optimize the classification model for environmental sound recognition.  The data samples have a length of up to 5 seconds. The MFCC feature extraction is utilized, and the mel-spectrograms are generated. The original images have a dimension of 768 × 384. To minimize the number of training parameters in the diffusion probabilistic model training process, the original images are resized to 128 × 128, where each image has 128 frames (columns) and 128 bands (rows).

\subsubsection{Hyperparameters Setting for Data Augmentation }
To enhance the stability of model training, we apply each data augmentation technique to the data samples in a stochastic manner described by the hyperparameter called $p$. Each data sample will undergo no more than two random transformations and at least one transformation. The specific setting is presented in Table 1. 

The city ambience noise is composed of three pieces of sound.  Each sound has a duration of 60 seconds. To augment the data sample with this noise, we randomly select a segment of the noise that has the same length as the data sample. Then, we superimpose this segment on the original data sample by adding their amplitudes. This creates a new data sample that contains both the original signal and the city ambience noise.

Pitch shifting refers to altering a sound's pitch by increasing or decreasing its frequency. The pitch shift factor can quantify the degree of pitch shifting, which is defined as the ratio between the output and input frequencies. For instance, a pitch shift factor of 2 indicates that the output frequency is double the input frequency. Pitch shifting can be performed in two directions: up pitch shifting, which increases the frequency and raises the pitch; and down pitch shifting, which decreases the frequency and lowers the pitch. The transformation of up and down pitch shifting will not be implemented simultaneously.

 Time stretch is changing the speed or duration of an audio signal without affecting its pitch. The minimum rate and maximum rate parameters control the minimum and maximum speed change factor, respectively. For example, a minimum rate of 0.8 means 20\% can slow the audio down, and a maximum rate of 1.25 means the audio can be sped up by 25\%.

\begin{table}[h]
\caption{Parameters Setting for Traditional Data Augmentation}
\centering
\begin{tabular}{ p{5.4cm}  p{2cm}   p{4cm}  }

\hline
Method&  $p$ & Setting \\
\hline

   Noise from City Ambience & 0.6 &The weight of the ambience is 0.6.\\
   Up   Pitch Shifting &0.8   &  The pitch shift factor is 2.\\
   Down Pitch Shifting  &0.8  & The pitch shift factor is 2.\\
    Time stretch &  0.7    &  Minimum rate is 0.8 and Maximum rate is 1.25 \\
\hline

\end{tabular}
\label{tab1}
\end{table}

\subsubsection{Diffusion Probabilistic Model Training Setting}

In this paper, we use the U-net  structure implemented in the \cite{b_29_pytorch_diffusion} for the estimation of ${\mathbf{p}}_{\theta}$. U-net performs four upsampling operations in its decoder path and uses skip connections to concatenate feature maps from the same stage in the encoder path instead of directly supervising and backpropagating loss on high-level semantic features\cite{b_30_unet}. This ensures that the final recovered feature map integrates more low-level features that capture fine details and enables the fusion of features at different scales for multi-scale prediction. The four upsampling operations also make the image recover edge information more finely by reducing spatial resolution loss.  Consequently, the U-net structure suffers much less information loss compared with other methods when sampling. As a result, the U-net structure is preferred in the diffusion probabilistic model.

The Unet architecture was configured with the following parameters: the number of channels in the first convolutional layer was set to 64 (dim=64), the number of channels was multiplied by 1, 2, 4, and 8 in the subsequent downsampling and upsampling layers (dim\_mults=(1,2,4,8)), the number of residual blocks in each group was fixed at 8 (resnet\_block\_groups=8), and the dimensionality of the learned sinusoidal embeddings was chosen as 16 \\(learned\_sinusoidal\_dim=16).

The optimization algorithm used for training the U-Net network is AdamW \cite{b_34_adamw}, a variant of Adam \cite{b_35_adam} incorporating weight decay regularization. The learning rate is set to 0.0001, a standard value for deep learning models. The weight decay parameter is set to 0.05, which helps to prevent overfitting by penalizing large weights. The other parameters of AdamW, such as beta values and epsilon values, are kept at their default values as suggested by the original paper \cite{b_34_adamw}. 

The loss function used to measure the discrepancy between the predicted and ground truth images is the mean squared error (MSE), defined as the average of the squared differences between the pixel values. The MSE is a widely used loss function for image reconstruction tasks, as it encourages high-fidelity reconstructions. 

The number of training epochs is set to 3500 for the dataset with augmented data, which is sufficient for the network to converge to a stable solution.  

To demonstrate the effectiveness of our network, we refrain from using techniques such as model transfer, exponential moving average (EMA), pretraining, or other tricks.

\subsubsection{Diffusion Probabilistic Model Sampling Setting}

 There are two versions of DPM-Solver++: one is DPM-Solver$++$ (2S), which is a second-order single-step solver, and the other is DPM-Solver$++$ (2M), which is a multistep second-order solver. The latter deals with the instability problem of the high-order solvers by reducing the effective step size. In our paper, we use the 2M as the sampling schedule in our study. The method is implemented via diffusers\cite{b_43_diffusers}.

The following settings are used for the DPM-Solver$++$ parameters: 

The initial and final values of $\beta$ for inference are 0.0001 and 0.02, respectively. $\beta$ is a hyperparameter that regulates the balance between the data likelihood and the prior distribution over the latent variables. The latent variables are unobserved variables that capture the underlying structure of the data. 

We use the linear method for the $\beta$ schedule that maps a range of betas to a sequence of betas for updating the model. The linear method progressively increases $\beta$ from the initial value to the final value over a predetermined number of iterations. The order of the DPM-Solver$++$ is 2, which indicates that it employs a second-order differential equation to model the dynamics of the latent variables. 

The solver type for the second-order solver is the midpoint, which is a numerical method that approximates the solution of the differential equation by using the midpoint of an interval as an estimate of its values.

The number of inference steps is 20. The parameter means the number of diffusion steps used when generating samples with a pre-trained model.

Other hyperparameters are set as default.

\subsubsection{Classification Models' Training Setting}

We trained our model for 500 epochs with a batch size of 30. We used the AdamW optimizer with the same hyperparameters as the DPMs training optimizer, such as learning rate, weight decay and epsilon. We used the cross-entropy loss function with label smoothing of 0.1 to prevent overfitting and improve generalization. We kept the other hyperparameters as default in the timm, such as dropout rate, hidden size, the number of layers, etc. The hyperparameter $k$ of top-k selection is set as one in the following experiments. The experiments are performed on a computer with a 13th Gen lntel R CoreTM i9-13900KF CPU and a GeForce RTX 4090 GPU (24 GB GDDR6X
memory).


\subsubsection{Discriminator Training Setting}

This paper employs the Xception model implemented by timm \cite{b_32_timm} as the backbone of the proposed method. The Xception model is a deep convolutional neural network that  can achieve high accuracy on various image recognition tasks. The following parameters are used to configure the Xception model for this paper:

We used the Xception model with the following hyperparameters: number of input channels = 3, dropout rate = 0., and global pooling method = average pooling.

We used the same training hyperparameters as the classification models, such as learning rate, batch size, number of epochs and optimizer.

\subsection{Experiments Results}

\subsubsection{Random Visual Samples of Generated Data}Fig. 2 illustrates random visual examples of the generated
spectrograms using DPMs. As you can see in this figure,
DPMs have a high capability to produce spectrograms that have
similar structures.

\begin{figure}
\centering
\centerline{\includegraphics[width=1.\columnwidth]{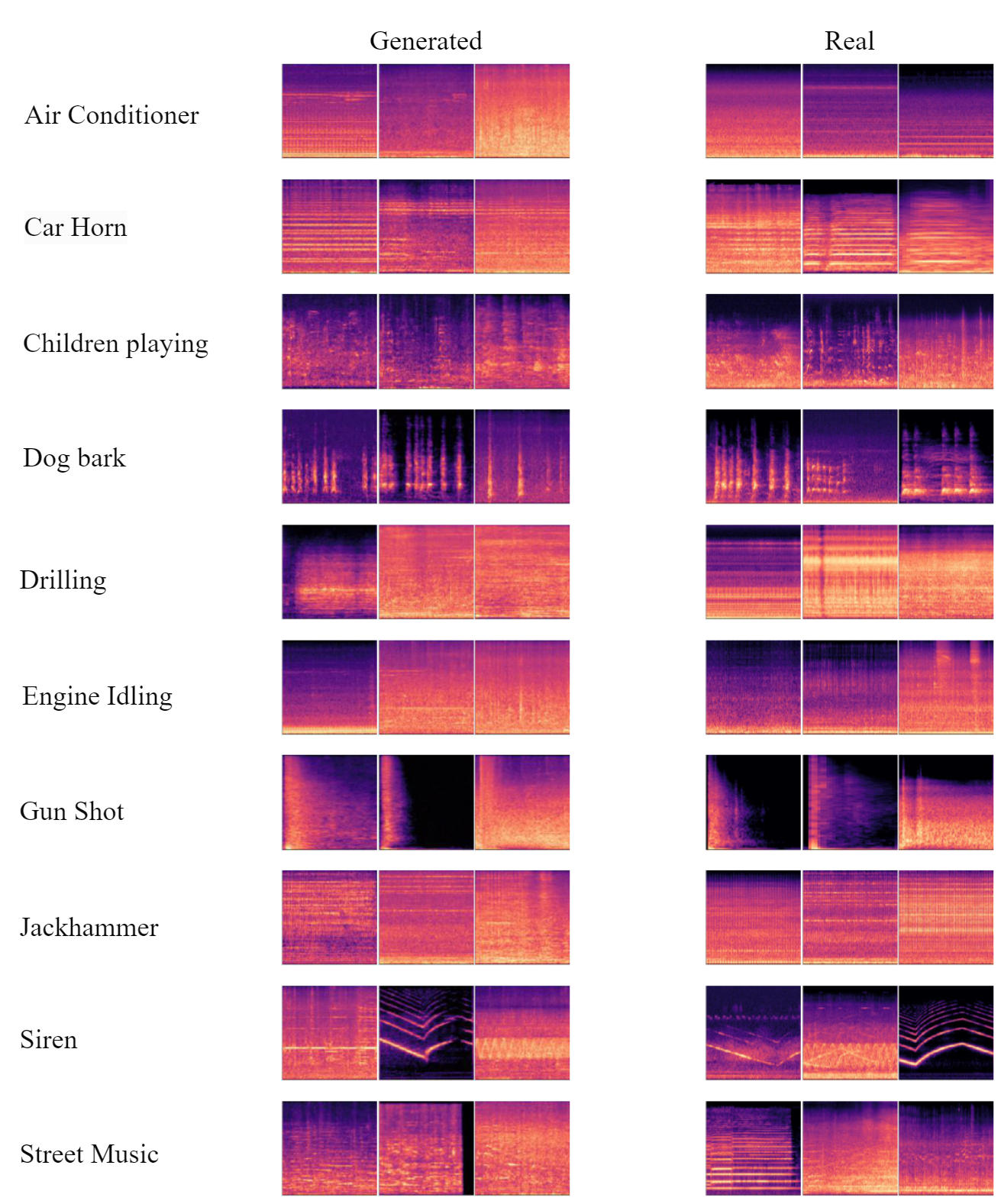}}
\caption{\quad Real (right panel) and generated (left panel) audio samples intelligent augmentation. Each row represents one sound class.}
\end{figure}

\subsubsection{Results of Augmentation for Different SOTA Models}

To assess the testing performance of the DL models and their generalization ability on unseen data, we use the 10-fold method as a fair and reliable evaluation technique. The final accuracy is computed as the mean of the 10 test results.

Table 2 shows the testing performance of seven DL models, ResNet-50 \cite{b_28_ResNet}, Xception \cite{b_38_inception}, ConViT-tiny \cite{b_39_convit}, Mobilevitv2-50 \cite{b_40_mobile}, Mobilevitv2-150 \cite{b_40_mobile}, Conv\\Next-tiny \cite{b_41_conv} and Deit III \cite{b_42_deit}, for environmental sound classification on UrbanSound8K dataset with synthetic augmentation + real dataset and with traditional data augmentation + real dataset.

It is evident that considerable improvements have been achieved
for all DL models by incorporating expanded datasets
with synthetic images, in terms of more stable training
(see loss curves in Fig 3) and higher accuracy and lower losses. For example, with the synthetic augmented
images, Inception-v3 and ResNet-50 yield classification
accuracies of 80.1\% and 80.2\% on the UrbanSound8K dataset, respectively representing about 6.3\%
and 7.6\% improvements over the baseline models without synthetic augmentation.

\begin{table}
\caption{Performance comparison of DL models on environmental sound classification trained with and without the samples generated
with the proposed data augmentation. The best values are in bold.}
\centering
\begin{tblr}{
  cell{1}{1} = {c=2,r=2}{},
  cell{1}{3} = {r=2}{},
  cell{1}{4} = {r=2}{},
  cell{4}{1} = {c=2}{},
  cell{6}{1} = {c=2}{},
  cell{11}{1} = {c=2}{},
  cell{13}{1} = {c=2}{},
  hline{1} = {-}{},
  hline{2} = {5}{},
  hline{3,10,17} = {-}{},
}
Index            &   & Models          & Parameters & UrbanSound8K       \\
                 &   &                 &            & Top-1 Accuracy (\%) \\
~                & ~ & ResNet-50       & 23528522   & \ \ \ \ \ \ \ \  73.8\%              \\
~                &   & Xception        & 20827442   & \ \ \ \ \ \ \ \  72.5\%              \\
~                & ~ & ConViT-tiny     & 5494098    & \ \ \ \ \ \ \ \  65.1\%              \\
Real+Traditional &   & Mobilevitv2-50  & 1116163    & \ \ \ \ \ \ \ \  59.2\%              \\
~                & ~ & Mobilevitv2-150 & 9833443    & \ \ \ \ \ \ \ \  68.3\%              \\
~                & ~ & ConvNext-tiny   & 27827818   & \ \ \ \ \ \ \ \  60.7\%              \\
~                & ~ & Deit
  III      & 85722634   &\ \ \ \ \ \ \ \  67.4\%              \\
~                & ~ & ResNet-50       & 23528522   & \ \ \ \ \ \ \ \  \textbf{80.1\% }    \\
~                &   & Xception        & 20827442   & \ \ \ \ \ \ \ \  \textbf{80.2\% }    \\
~                & ~ & ConViT-tiny     & 5494098    & \ \ \ \ \ \ \ \  \textbf{68.6\% }    \\
Real+Synthetic   &   & Mobilevitv2-50  & 1116163    & \ \ \ \ \ \ \ \  \textbf{62.0\% }    \\
~                & ~ & Mobilevitv2-150 & 9833443    & \ \ \ \ \ \ \ \  \textbf{74.6\% }    \\
~                & ~ & ConvNext-tiny   & 27827818   & \ \ \ \ \ \ \ \  \textbf{65.9\% }    \\
~                & ~ & Deit
  III      & 85722634   &\ \ \ \ \ \ \ \  \textbf{72.2\% }    
\end{tblr}
\end{table}


\begin{figure}
\label{Loss}
\centering
\centerline{\includegraphics[width=1.1\columnwidth]{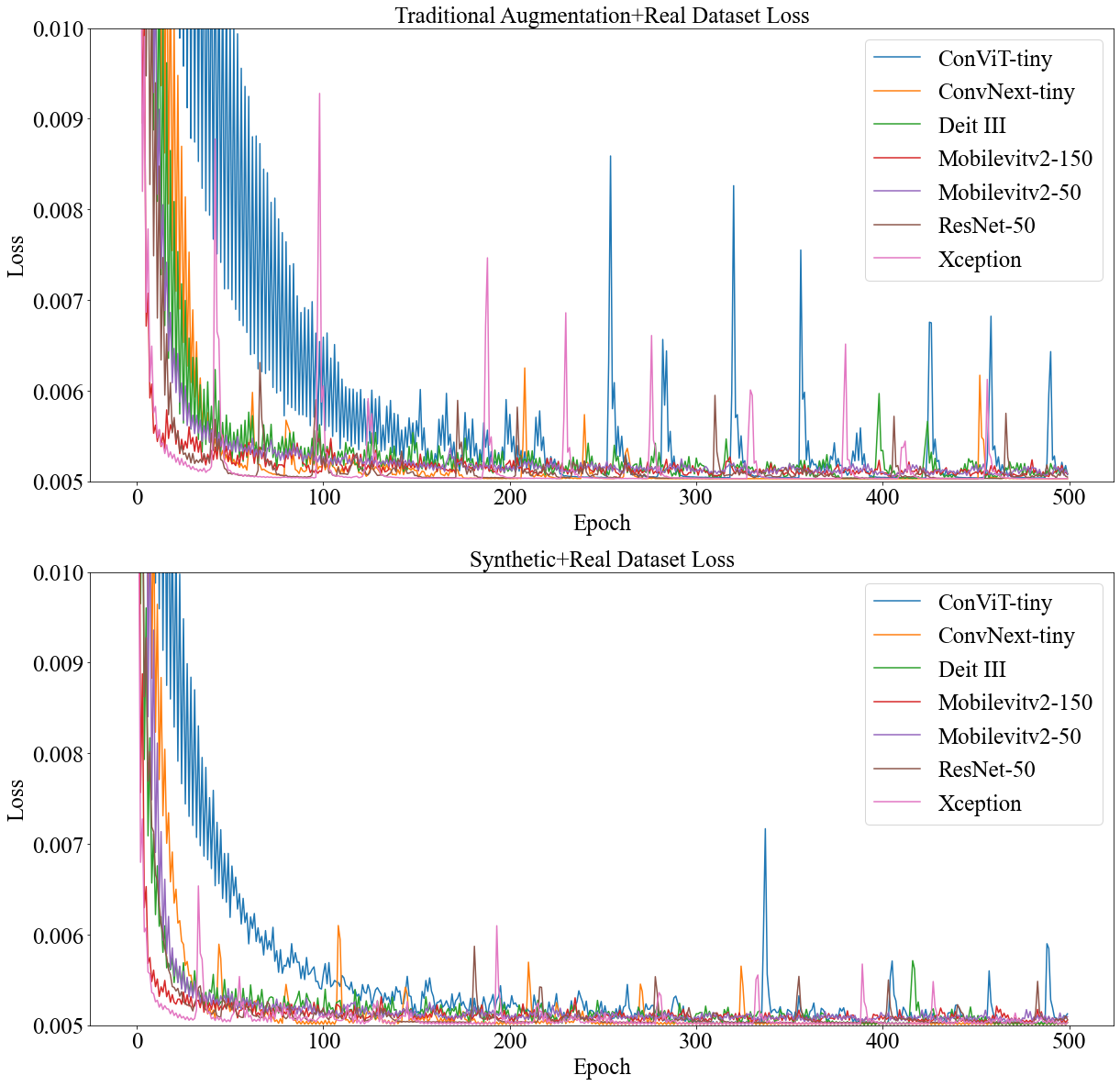}}
\caption{\quad Training loss curves of different DL models on the UrbanSound8K dataset with and without synthetic data augmentation. The left figure represents the baseline models trained without synthetic augmentation, while the right figure represents those trained with synthetic augmentation. The results show that synthetic augmentation helps to reduce the loss and improve the stability of the training process for all models.}
\end{figure}

\subsubsection{Influence of Hyperparameter $k$ for Top-k Selection}

To assess how the hyperparameter $k$ affects the outcome of synthetic data generation, we perform a series of experiments with different values of $k$ from 1 to 10 and compare the results using accuracy. The results are shown in Table 4.

\begin{table}
\centering
\caption{The impact of hyperparameter k on the accuracy of seven deep learning models on a synthetic+real dataset. The table shows how the accuracy changes as k varies from 1 to 10. The value of $k$ does not affect the generation of data samples when $k>=5$ in our training setting. Therefore, the accuracy of the models is constant for these values of $k$. The best values are in bold.}
\begin{tblr}{
  cell{1}{1} = {r=2}{},
  cell{1}{2} = {c=5}{},
  hline{1} = {-}{},
  hline{2} = {2-6}{},
  hline{3,10} = {-}{},
}
Method ~~~      & \ \ \  \ \ \  \ Different
  Values of Hyperparameter k &                  &        &                  &                  \\
                & 1                                      & 2                & 3      & 4                & $\geq$5               \\
ResNet-50       & \textbf{80.1\% }                       & 80.2\%           & 77.3\% & 76.7\%           & 77.4\%           \\
Xception        & \textbf{80.2\% }                       & 78.7\%           & 74.8\% & 74.5\%           & 74.0\%           \\
ConViT-tiny     & 68.6\%                                 & \textbf{68.7\% } & 64.5\% & 64.5\%           & 64.6\%           \\
Mobilevitv2-50  & 62.0\%                                 & 62.8\%           & 63.0\% & 65.5\%           & \textbf{69.0\% } \\
Mobilevitv2-150 & 74.6\%                                 & 73.5\%           & 73.5\% & \textbf{76.4\% } & 72.6\%           \\
ConvNext-tiny   & \textbf{65.9\% }                       & 62.3\%           & 62.0\% & 61.0\%           & 62.8\%           \\
Deit III        & 72.2\%                                 & \textbf{72.8\% } & 71.3\% & 69.2\%           & 67.8\%           
\end{tblr}
\end{table}

The results show that the top-k selection strategy effectively enhances most DL models' performance. By selecting the most confident synthetic images for each class, the top-k strategy reduces the noise and ambiguity in the augmented data. As shown in Table 4, the top-k strategy significantly improves accuracy for six models. For instance, ResNet-50 achieves an accuracy of 80.1\% with k=1, which is 2.7\% higher than the model without top-k selection. Similarly, Xception attains an accuracy of 80.2\% with k=1, which is 6.2\% higher than the models without top-k selection.

\section{Conclusion}

This paper introduced a novel application of diffusion models for generating high-quality synthetic images from environmental sound recordings. To the best of our knowledge, this is the first study that explores the use of diffusion models for data augmentation in environmental sound classification. We also proposed a new selection method based on the top-k confidence scores to filter out the low-quality synthetic images and retain the most informative ones for each sound class. 

We conducted extensive experiments on a widely used sound dataset, UrbanSound8K, and evaluated the performance of seven state-of-the-art DL models trained on the augmented datasets with different settings. The experimental results demonstrated that the diffusion models can generate realistic and diverse synthetic images that can effectively improve the classification accuracy and reduce the losses for all DL models. Moreover, the top-k selection method further enhanced the performance by removing noisy and ambiguous synthetic images and increasing the data balance among classes.

\end{document}